\documentclass[aps,prb,reprint,a4paper,floatfix,groupedaddress,amsmath,amssymb,amsfonts]{revtex4-1}
\pdfoutput=1
\usepackage{mathptmx}
\usepackage{textcomp}
\usepackage{helvet}
\usepackage[utf8]{inputenc} 
\usepackage{graphicx}
\usepackage{upgreek} 
\usepackage{xspace}
\usepackage{textcomp}
\usepackage[pdftex]{hyperref}
\hypersetup{
	pdftitle={Crystal-phase quantum dots in GaN quantum wires},
	colorlinks,
	citecolor=blue
	}
\usepackage[rightcaption]{sidecap}
\usepackage{color,soul} 
\usepackage[
,textwidth=17.5cm
,textheight=23.5cm
,verbose
,dvips
]{geometry}

\usepackage[colorinlistoftodos,prependcaption,textsize=footnotesize]{todonotes}

\newcommand{\dox}{$(D^{0},X_A)$\xspace}
\newcommand{\sfx}{$(I_{1},X)$\xspace}

\begin{document}
\title{Crystal-phase quantum dots in GaN quantum wires}

\author{Pierre Corfdir}
\email{corfdir@pdi-berlin.de}
\author{Christian Hauswald}
\author{Oliver Marquardt}
\author{Timur Flissikowski}
\author{Johannes K. Zettler}
\author{Sergio Fernández-Garrido}
\author{Lutz Geelhaar}
\author{Holger T. Grahn}
\author{Oliver Brandt}

\affiliation{Paul-Drude-Institut für Festkörperelektronik,
Hausvogteiplatz 5--7, 10117 Berlin, Germany}


\begin{abstract}

We study the nature of excitons bound to I$_1$ basal plane stacking faults in ensembles of ultrathin GaN nanowires by continuous-wave and time-resolved photoluminescence spectroscopy. These ultrathin nanowires, obtained by the thermal decomposition of spontaneously formed GaN nanowire ensembles, are tapered and have tip diameters down to 6~nm. With decreasing nanowire diameter, we observe a strong blue shift of the transition originating from the radiative decay of stacking fault-bound excitons. Moreover, the radiative lifetime of this transition in the ultrathin nanowires is independent of temperature up to 60~K and significantly longer than that of the corresponding transition in as-grown nanowires. These findings reveal a zero-dimensional character of the confined exciton state and thus demonstrate that I$_1$ stacking faults in ultrathin nanowires act as genuine quantum dots.

\end{abstract}

\maketitle

Spontaneously formed GaN nanowires are comparable in structural perfection to state-of-the-art freestanding GaN.\cite{Zettler2015a} The nanowire geometry inhibits the propagation of threading dislocations along the nanowire axis, resulting in dislocation-free crystals regardless of the substrate.\cite{Hersee2011,Kishino2015} In contrast to group-III arsenide and phosphide nanowires, which are synthesized by vapor-liquid-solid growth and are prone to a pronounced polytypism,\cite{Bao2008,Corfdir2013} spontaneously formed GaN nanowires exclusively crystallize in the wurtzite lattice structure with only occasional I$_1$ basal plane stacking faults (BSFs).\cite{Calleja2007,Geelhaar2011} Consequently, the radiative transitions related to excitons bound to I$_1$ BSFs [$(I_{1},X)$] in GaN nanowires are spectrally well resolved and distinct from other excitonic transitions in GaN. This fact has been essential for shedding light on the nature of the \sfx.\cite{Corfdir2014b,Korona2014} In particular, for nanowires with a diameter larger than 50~nm, the \sfx was shown to exhibit a two-dimensional density of states, i.\,e., I$_1$ BSFs indeed act as quantum wells.\cite{Stampfl1998,Corfdir2014b} 

These so-called crystal-phase quantum structures are free of strain and alloy disorder, their interfaces are atomically abrupt.\cite{Akopian2010} In GaN nanowires, the decay of the \sfx is purely radiative up to 60~K.\cite{Corfdir2014b,Korona2014} BSFs thus form an exceptionally well-defined model system for fundamental studies of confined excitons. In this context, the recent fabrication of ultrathin GaN nanowires with a diameter down to 6~nm is of great interest.\cite{Zettler2016} Due to the mismatch in dielectric constants between GaN and air, excitons in these ultrathin nanowires experience a strong radial confinement, i.\,e., the ultrathin GaN nanowires act as quantum wires despite the fact that their diameter still exceeds at least twice the exciton Bohr radius.\cite{Zettler2016} In addition, the thermal decomposition technique used for the controlled thinning of the as-grown nanowires does not affect their high structural perfection. Finally, despite their extremely small diameter, these nanostructures exhibit a high radiative efficiency due to a rather slow surface recombination velocity at the nanowires' sidewall facets less than than 210~cm/s at 60~K.\cite{Corfdir2014b}

\begin{figure}
\includegraphics[scale=1.0]{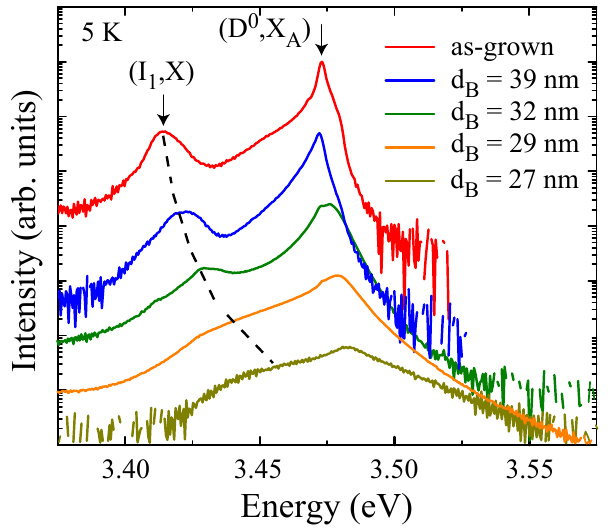}
\caption{(color online) PL spectra of the as-grown nanowire and the quantum wire ensembles at 5~K acquired with an excitation density of 10 mW/cm$^2$(the spectra have been shifted vertically). The average $d_B$ is specified for each sample. The dashed line is a guide to the eye highlighting the blueshift of the \sfx transition with decreasing $d_B$.}
\label{fig:FigureCWPL}
\end{figure} 

In this work, we use continuous-wave (cw) and time-resolved (TR) photoluminescence (PL) spectroscopy to investigate the radiative decay and the dynamics of the \sfx in GaN quantum wire ensembles fabricated by partial thermal decomposition. We demonstrate that I$_1$ BSFs in these nanowires act as quantum dots. With decreasing nanowire diameter, the \sfx transition blueshifts as a result of radial confinement. The radiative lifetime of the \sfx in crystal-phase quantum dots does not vary with temperature and is significantly longer than the one measured at 5~K for as-grown nanowires. Using self-consistent eight-band $\mathbf{k \cdot p}$ calculations, we show that the increase in radiative lifetime with decreasing diameter results from the reduced coherence area of the \sfx.

\begin{SCfigure*}
\includegraphics[scale=1.0]{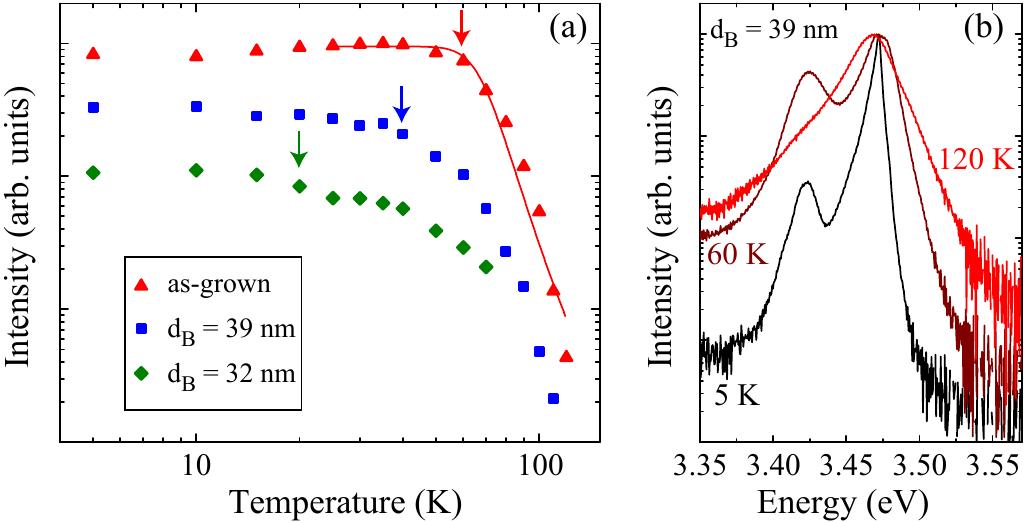}
\caption{(color online) (a) \sfx intensity $I$ as a function of temperature $T$ for the as-grown nanowire (triangles) and for the quantum wire ensembles with two different average $d_B$. The arrows show the temperature at which the emission intensity starts to quench. The solid line shows the best fit to the \sfx intensity using $I(T) \propto 1/ \left(1+a T \exp{\left[- \frac{E_a}{kT} \right] } \right)$, with $a$ a fitting parameter and $E_a = 57\pm 5$~meV the activation energy for the thermal escape of the \sfx from the BSF. (b) PL spectra of the quantum wire ensembles with $d_B = 39$~nm at 5, 60 and 120~K taken with an excitation density of 10 mW/cm$^2$ (the spectra have been normalized). }
\label{fig:FigurePLwithT}
\end{SCfigure*} 

Ultrathin GaN nanowires with a length of about 1~\textmu m have been obtained by partial thermal decomposition at 920\,°C of GaN nanowire ensembles formed during molecular beam epitaxy on a Si(111) substrate.\cite{Zettler2016} The length and the diameter of the as-grown nanowires are 2~\textmu m and 51~nm, respectively, and we estimate the density of I$_1$ BSFs to be about 1 per nanowire (see the cathodoluminescence mappings in Ref.~\onlinecite{Corfdir2014b}). The concurrent layer-by-layer desorption of atoms from the top surface and from the sidewalls leads to tapering. The thinnest nanowires obtained exhibit an average base diameter $d_B = 27$~nm while the diameter at their tip can be as small as 6~nm.\cite{Zettler2016} Following the results in Ref.~\onlinecite{Zettler2016}, ultrathin nanowires are referred to as quantum wires in the remainder of the paper. Charging effects leads to an overestimation of the tip diameters when measuring such thin wires by top-view and cross-sectional secondary electron microscopy.\cite{Zettler2016,Loitsch2015} Therefore, the emission properties of GaN quantum wires were correlated with their average $d_B$. Note, however, that the emission is very likely to originate from sections of the nanowires with diameter substantially smaller than $d_B$. 

Continuous-wave (cw) PL experiments were performed using the 325~nm line of a HeCd laser for excitation. The laser was focused onto the sample to a diameter of 60~µm. The PL signal was analyzed using a monochromator followed by a charge-coupled device camera for detection. Time-resolved (TR) PL spectroscopy was carried out using the second harmonic of fs pulses obtained from an optical parametric oscillator pumped by a Ti:sapphire laser (emission wavelength and repetition rate of 325~nm and 76~MHz, respectively). The energy fluence per pulse was kept below 0.3~\textmu J/cm$^2$. The transient emission was spectrally dispersed by a monochromator and detected by a streak camera operating in single shot mode.  For both cw and TR PL measurements, the samples were mounted in a coldfinger cryostat whose temperature can be varied between 5 and 300~K. For both experiments, the laser was polarized perpendicular to the nanowire axis. As the average nanowire diameter for all of samples is well within the sub-wavelength range, the coupling of light into the nanowires is getting less and less efficient with decreasing diameter, thus strongly reducing absorption. The photogenerated carrier density in the quantum wires should thus be \emph{lower} as compared to the as-grown nanowires.

Figure \ref{fig:FigureCWPL} shows the PL spectra at 5~K for an ensemble of as-grown nanowires and for ensembles of partially decomposed, quantum wires with $d_B$ between 39 and 27~nm. The spectrum for the as-grown ensemble is dominated by the recombination of A excitons bound to neutral O donors at 3.471~eV [$(D^{0},X_A)$]. The lower energy band centered at 3.410~eV is related to the recombination of excitons bound to I$_1$ BSFs.\cite{Corfdir2014b,Korona2014}

Decreasing $d_B$ from 51 to 27~nm, the energy of the \dox transition increases from 3.471 to 3.481~eV (Fig.~\ref{fig:FigureCWPL}), indicating a progressively stronger confinement of the excitons in the corresponding nanowires. This confinement is caused by the mismatch in dielectric constants between GaN and vacuum at the nanowire sidewalls.\cite{Keldysh1979,Kumagai1989,Zettler2016} An even stronger blueshift (42~meV) with decreasing nanowire diameter is observed for the \sfx line, suggesting that the \sfx state in GaN quantum wires is radially confined. In other words, $I_1$ BSFs in GaN quantum wires seem to act as crystal-phase quantum dots. The  larger blueshift of the \sfx as compared to the one observed for the \dox line probably arises from the different location of the exciton states involved in these transitions: whereas donors are distributed uniformly along the entire length of the nanowire, I$_1$ BSFs may be located preferentially in the top parts of the nanowires, where the diameter is smaller and the confinement of the exciton stronger. This result is consistent with the fact that BSFs result from the nanowire coalescence and form several hundreds of nm above the contact point between adjacent nanowires.\cite{Consonni2009} We also observe that the thinnest nanowires exhibit the broadest \sfx and \dox lines at 5~K (Fig.~\ref{fig:FigureCWPL}). This finding is a direct consequence of the increase in confinement with decreasing diameter: the thinner the nanowire, the larger the impact of diameter fluctuations on the \dox and \sfx energies and hence the larger the broadening of the corresponding emission lines. As a result of the significant spectral overlap between the \sfx and \dox transitions for the thinnest nanowires (Fig.~\ref{fig:FigureCWPL}), we focus in the following on the optical properties of the quantum wires with $d_B = 39$ and 32~nm.

\begin{figure*}
\includegraphics[scale=1.0]{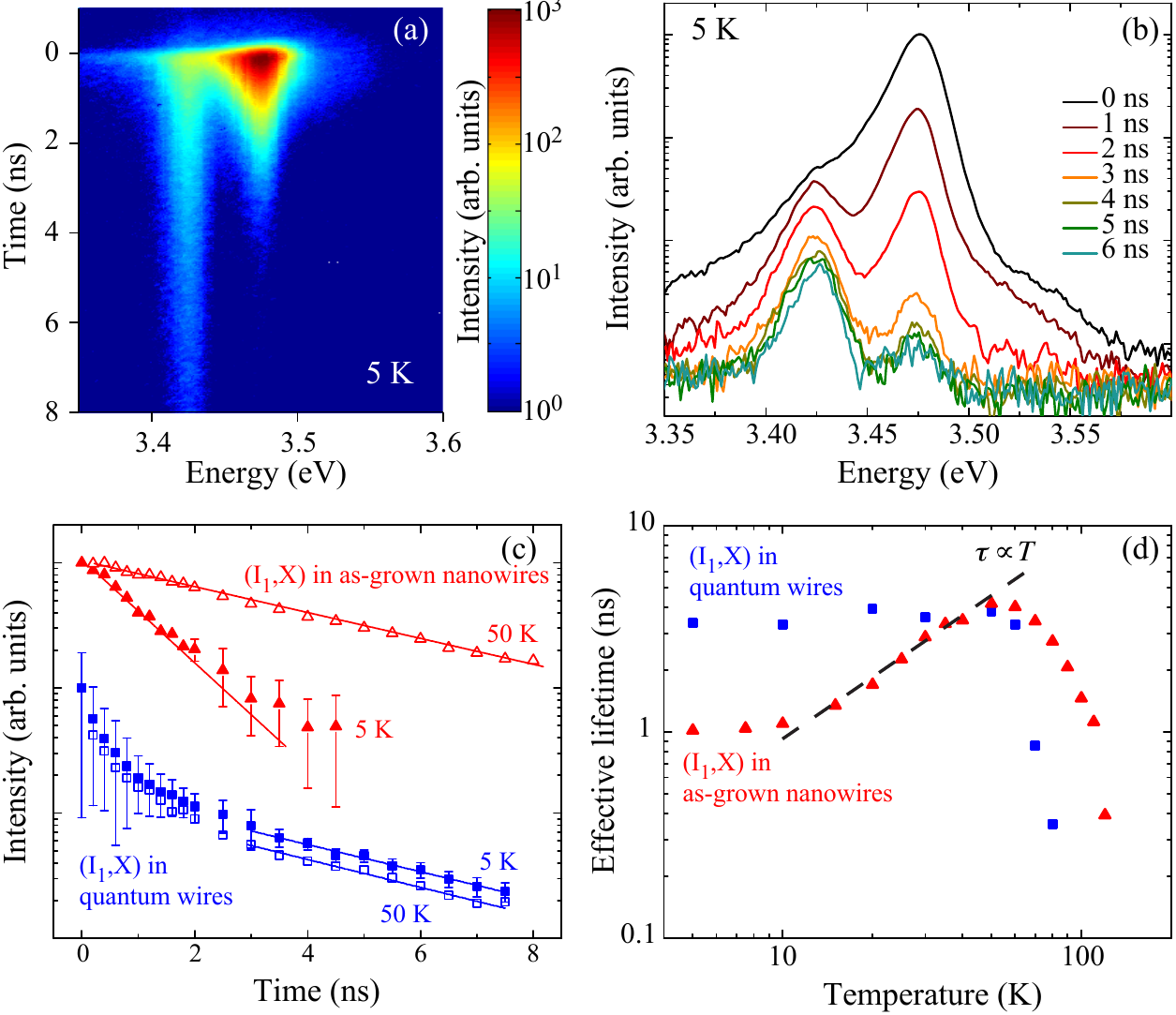}
\caption{(color online) (a) Streak camera image of an ensemble of quantum wires with $d_B = 39$~nm recorded at 5~K with an energy fluence per pulse of 0.3~\textmu J/cm$^2$. The intensity is displayed on a logarithmic scale from blue (low intensity) to red (high intensity). (b) Temporal evolution of the PL spectra of an ensemble of quantum wires with $d_B = 39$~nm at 5~K. (c) PL intensity transients of the \sfx transition at 5 and 50~K (solid and open symbols, respectively) for the ensembles of as-grown nanowires (triangles) and quantum wires with $d_B = 39$~nm (squares). The peak intensities are normalized. The transients for the quantum wires have been shifted vertically for clarity. The solid lines are exponential fits of the transients. (d) Temperature dependence of $\tau$ for the ensembles of as-grown nanowires (triangles) and quantum wires with $d_B = 39$~nm (squares). The lifetimes $\tau$ have been obtained from the single exponential fits shown in (c). The dashed line is a guide to the eye showing the linear increase of $\tau$ for the as-grown nanowires between 15 and 60~K.}
\label{fig:FigureTRPL}
\end{figure*} 

Figure~\ref{fig:FigurePLwithT}(a) shows the temperature dependence of the integrated intensity of the \sfx line for the as-grown nanowire ensembles and for two quantum wire ensembles with $d_B = 39$ and 32~nm. For the as-grown nanowire ensemble, the intensity of the \sfx transition remains constant between 4 and 60~K, indicating that the recombination of the \sfx is purely radiative up to 60~K.\cite{Corfdir2014b} For temperatures above 60~K, excitons can thermally escape from the comparatively shallow crystal-phase quantum well, leading to an abrupt quenching of the \sfx line.\cite{Graham2013,Rudolph2013} The intensity of the \sfx transition follows an Arrhenius behavior with an activation energy $E_a = (57 \pm 5)$~meV [Fig.~\ref{fig:FigureCWPL}(b)], coinciding with the energy difference between the \sfx and the free exciton in fault-free segments.\cite{Rudolph2013,Corfdir2014b} For the ensembles of quantum wires, the \sfx transition also shows a constant PL intensity in the low temperature range before decreasing strongly at high temperatures [Fig.~\ref{fig:FigurePLwithT}(a)]. Figure~\ref{fig:FigurePLwithT}(b) shows normalized PL spectra taken at 5, 60 and 120~K on the sample with $d_B = 39$~nm. The intensity ratio between the \sfx and the free exciton decreases strongly between 60 and 120~K, indicating that the quenching of the \sfx PL in the quantum wires is also due to the thermal escape of the exciton from the BSFs. The range of constant PL intensity however decreases with decreasing $d_B$, and the intensity quenching becomes less abrupt than for the as-grown nanowires. Both of these findings are easily understood: since a smaller diameter results in a larger confinement energy for the \sfx state (Fig.~\ref{fig:FigureCWPL}), it also results in a lower value of $E_a$. Furthermore, the tapering of the quantum wires leads to a distribution of $E_a$ values, which manifests itself in a gradual quenching of the \sfx PL intensity as compared to that observed for the as-grown, non-tapered nanowire ensemble.

\begin{figure}
\includegraphics[scale=1.0]{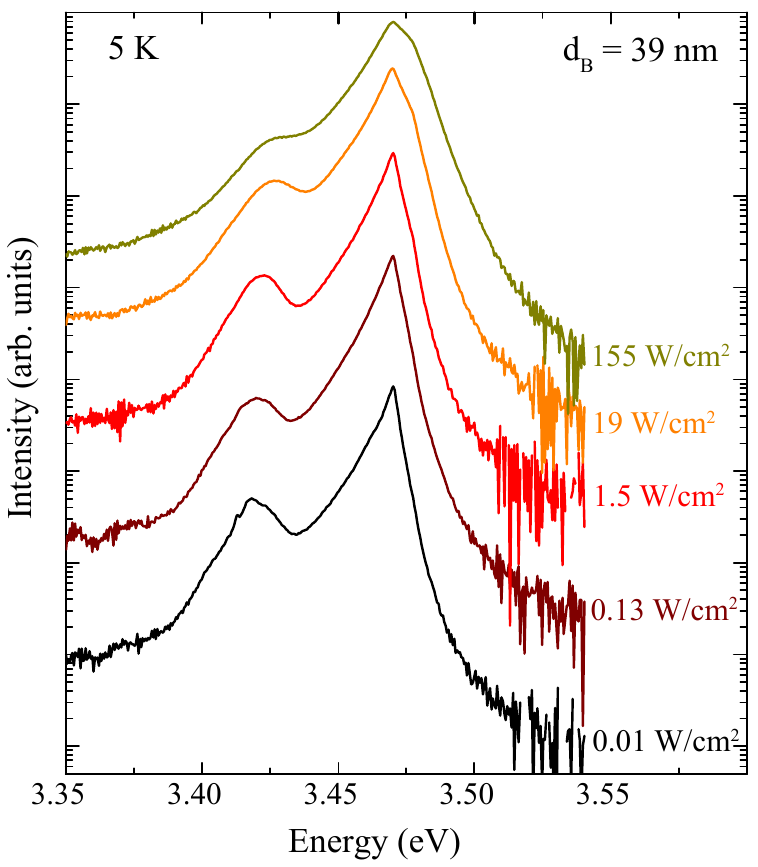}
\caption{(color online) Continuous-wave photoluminescence spectra of an ensemble of quantum wires with $d_B = 39$~nm at 5~K as a function of the excitation power density.}
\label{fig:FigurePowerSeries}
\end{figure} 

Figure~\ref{fig:FigureTRPL}(a) displays a streak camera image taken at 5~K on the sample with $d_B=39$~nm. Spectral profiles taken at various time delays are shown in Fig.~\ref{fig:FigureTRPL}(b). The \dox PL decays exponentially with a decay time of 390~ps. This fast decay has a nonradiative origin and most probably arises from exciton recombination at point defects.\cite{Hauswald2014} Figure~\ref{fig:FigureTRPL}(c) shows PL intensity transients of the \sfx line at 5 and 50~K for the as-grown nanowires and for the quantum wire ensemble with $d_B = 39$~nm. The integrated intensity of the \sfx line has been obtained by a spectral deconvolution of the transient spectra.\cite{Hauswald2013} Note that due to the significant spectral overlap between the \sfx and the \dox transitions at the early stage of the decay [Fig.~\ref{fig:FigureTRPL}(b)], the values obtained for the \sfx PL intensities of the partially decomposed nanowire sample exhibit a comparatively large uncertainty for the initial 1~ns. At 5~K, the \sfx state for the as-grown nanowires decays exponentially, and the decay time $\tau = 1.0$~ns is identical with the radiative lifetime $\tau_\text{r}$. The \sfx PL decay for the quantum wires with $d_B = 39$~nm is more complex: it is nonexponential during the first two ns after excitation, and becomes exponential with a decay time $\tau = 3.4$~ns thereafter. Comparable results have been obtained for the sample with $d_B = 32$~nm (not shown). Two different phenomena may, in principle, account for this initial nonexponential decay. 

First, the discontinuity of the polarization field at the interfaces of I$_1$ BSFs induces strong electrostatic fields along the nanowire axis, spatially separating the electron and hole wavefunctions.\cite{Sun2002,Laehnemann2012,Corfdir2012} A high initial carrier density created by pulsed excitation may screen these fields, giving rise to a minimum value for the radiative lifetime $\tau_\text{r}$ directly after excitation. Since the carrier density is subsequently reduced by recombination, the electric fields are restored again with time, resulting potentially in a continuous decrease in the \sfx energy as well as in a continuous increase in $\tau_\text{r}$.\cite{Reale2003,Bretagnon2006} Since the \sfx recombination is purely radiative at low temperature [Fig.~\ref{fig:FigurePLwithT}(a)], the latter increase could explain the nonexponential decay of the \sfx emission after pulsed excitation. However,  the \sfx energy remains constant during the whole decay [Fig.~\ref{fig:FigureTRPL}(b)]. This finding suggests that the change in carrier density with time after pulsed excitation does not lead to strong modifications in the strength of built-in electric fields, and that the nonexponential \sfx PL decay seen in Fig.~\ref{fig:FigureTRPL}(c) is not due to the dynamical descreening of these fields. To confirm this result, we have recorded excitation-density dependent cw PL spectra at 5~K  on the sample with $d_B = 39$~nm (Fig.~\ref{fig:FigurePowerSeries}). Increasing the excitation density from 0.01 to 1.5~W/cm$^2$ does not lead to any change in the energy of the \sfx, confirming that screening is negligible in this range of excitation densities.\cite{Reale2003} Note that the small blueshift observed for larger excitation densities most probably arises from band filling and/or heating effects (see the change in the \dox and free exciton lineshapes when the density increases from 1.5 to 155 W/cm$^2$). In view of the results in Figs.~\ref{fig:FigureTRPL}(b) and \ref{fig:FigurePowerSeries}, it is unlikely that the nonexponential decay observed for the \sfx in the quantum wires [Fig.~\ref{fig:FigureTRPL}(c)] originates from the dynamical descreening of the built-in electric fields. 

Second, the origin for the nonexponential decay may be associated with the pronounced tapering of the quantum wires. Since the BSFs are likely to occur at different positions along the nanowire axis, their radial dimension and thus the degree of radial confinement also varies. Since the radiative lifetime almost certainly depends on this degree of radial confinement,\cite{Kavokin1994,Bellessa1998} a multiexponential decay would be an inevitable consequence. Following this interpretation, the experimental result of longer decay times for the quantum wires implies that the radiative lifetime increases with decreasing diameter. We will return to this issue after a discussion of the transients at 50~K and the temperature dependence of $\tau$.

At a temperature of 50~K, for which the PL intensity of our samples is still close to that at 5~K [cf.\ Fig.~\ref{fig:FigurePLwithT}(a)], the decay of the \sfx remains exponential for the as-grown nanowires, but $\tau$ increases from 1.0 to 4.2~ns. In contrast, the increased temperature does not affect the decay of the \sfx for the quantum wires. Figure~\ref{fig:FigureTRPL}(d) shows the evolution of $\tau$ between 5 and 120~K for both samples. Up to a temperature of 50--60~K, the decay of the \sfx is purely radiative and $\tau=\tau_\text{r}$. Between 15 and 40~K, $\tau_\text{r}$ increases nearly linearly for the as-grown nanowires, demonstrating that (i) I$_1$ BSFs in nanowires with a diameter of 51~nm act as quantum wells,\cite{Corfdir2011} and (ii) the \sfx is free to move along the BSF plane. The deviation from a linear behavior for temperatures lower than 15~K arises from exciton localization along the BSF plane due to the presence of donors as discussed in Ref.~\onlinecite{Corfdir2014b}. The significant reduction in $\tau$ for temperatures larger than 60~K is due to the thermal escape of excitons from the BSF plane. For the quantum wires, the long component of the nonexponential decay is independent of temperature and corresponds to the radiative lifetime of the \sfx in the thinnest nanowire segments according to the discussion above. The constant lifetime confirms unambiguously that BSFs in GaN quantum wires behave as crystal-phase quantum dots. The reduction of the dimensionality of the \sfx state from two to zero already when $d_B = 39$~nm again suggests that the crystal-phase quantum dots are located in the top part of the nanowires, where the diameter is much smaller.

\begin{figure}
\includegraphics[scale=1.0]{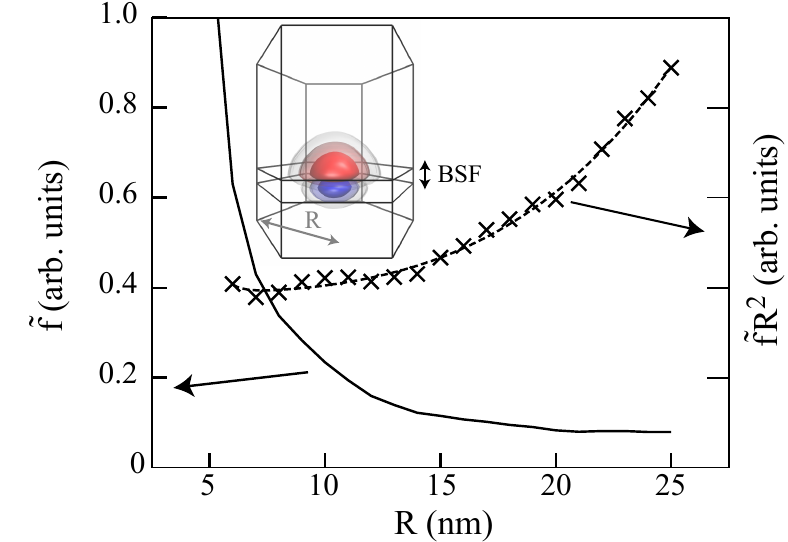}
\caption{(color online) Oscillator strength per unit area $\tilde{f}$ and total oscillator strength $\tilde{f} R^2$ for the \sfx as a function of the disk diameter (solid and dashed lines, respectively). The inset shows the electron (red) and hole (blue) ground state charge densities for a nanowire diameter of 20~nm. The inset has been prepared using Visual Molecular Dynamics.\cite{Humphrey1996}}
\label{fig:FigureKP}
\end{figure} 

Finally, we address the origin of the increase in $\tau_\text{r}$ with decreasing diameter observed above. Considering that the exciton is coherent over the entire BSF, the radiative decay rate  $\Gamma_r=1/\tau_r$ can be written approximately as:\cite{Brandt1992,Kavokin1994}

\begin{align}
\Gamma_r \propto \tilde{f} \, A_c \propto | \langle \chi_\text{e}(z) | \chi_\text{h}(z) \rangle |^2 \left( \frac{R}{a_\perp} \right)^2 
\label{eq1}
\end{align}  

with the oscillator strength per unit area $\tilde{f}$, the coherence area $A_c$ (which is assumed to be determined by the nanowire's radius $R$),\cite{Brandt1992,Kavokin1994,Bellessa1998} the overlap integral between the electron and hole wavefunctions along the nanowire axis $\langle \chi_\text{e}(z) | \chi_\text{h}(z) \rangle$, and the Bohr radius of the exciton in the BSF plane $a_\perp$. For examining the change of $\tau_\text{r}$ with a decrease in $R$, we calculate the wavefunction of the \sfx in nanowires of diameters between 10 and 50~nm using eight-band $\mathbf{k \cdot p}$ calculations.\cite{Marquardt2014} The $I_1$ BSF within the GaN nanowire is represented by 3 monolayers of zincblende GaN within a wurtzite GaN segment of 20~nm length. The nanowires are considered to be undoped, i.\,e., surface potentials are absent. The surface is considered as an infinite potential barrier. The spontaneous polarization of wurtzite GaN induces axial electrostatic fields in the $I_1$ BSF with a magnitude of 3~MV/cm.\cite{Corfdir2012} We average over all polarization directions, i.\,e., the oscillator strength $\tilde{f}$ is simply given by $ | \langle \chi_\text{e}(z) | \chi_\text{h}(z) \rangle |^2 /a_\perp^2$.
 
Figure~\ref{fig:FigureKP} shows the dependence of $\tilde{f}$ and $\tilde{f} R^2$ as a function of $R$. As shown in the inset of Fig.~\ref{fig:FigureKP} for a diameter of 20~nm, both the electron and the hole are located in the center of the quantum disk. With decreasing $R$, the electron experiences a progressively stronger confinement which results in a strong increase of $\tilde{f}$ as displayed in Fig.~\ref{fig:FigureKP}. A radial separation between the electron and hole wavefunctions as observed in Ref.~\onlinecite{Marquardt2014} does not occur for the range of nanowire diameters considered here, a finding that still holds when considering the presence of surface potentials due to Fermi level pinning at the free sidewalls and a homogeneous background doping of $10^{17}$~cm$^{-3}$ (not shown here). The increase in $\tilde{f}$ with decreasing $R$ should result in a decrease in $\tau_\text{r}$, in contradiction to our experimental observation [cf.\ Fig.~\ref{fig:FigureTRPL}(c,d)].

As shown in Fig.~\ref{fig:FigureKP}, the dependence of the radiative decay rate on diameter is reversed when taking into account the factor $R^2$ in Eq.~(\ref{eq1}). This factor accounts for the fact that the exciton's radiative decay is enhanced by its coherent macroscopic polarization.\cite{Brandt1992,Kavokin1994,Bellessa1998} This enhanced radiative decay may also be understood in the context of the arguments of \citet{Rashba1962}: the larger the coherence area of the exciton, the smaller the spread of its wavefunction in \textbf{k} space and thus the shorter $\tau_\text{r}$. As shown in Fig.~\ref{fig:FigureKP}, the total oscillator strength $\tilde{f} R^2$ is indeed predicted to decrease with decreasing $R$, and the radiative lifetime $\tau_\text{r}$ is thus expected to increase correspondingly in agreement with the experiment. 

To conclude, changing the diameter of GaN nanowires in a controlled fashion has allowed us to observe the transition from two- to zero-dimensional stacking-fault bound excitons. This transition occurs at diameters significantly larger than the exciton's Bohr radius and is induced by dielectric confinement. Because of the absence of structural imperfections such as interfacial steps and alloy fluctuations, the radiative decay rate of these excitons scales with the nanowire diameter, which imposes a geometrical limitation of their coherence area.    

We thank Alberto Hernández-Mínguez for carefully reading our manuscript. Partial funding from the Deutsche Forschungsgemeinschaft within SFB 951 is gratefully acknowledged. P.C. acknowledges partial funding from the Fonds National Suisse de la Recherche Scientifique through project 161032.

\bibliography{bibliography}

\begin{thebibliography}{32}%
\makeatletter
\providecommand \@ifxundefined [1]{%
 \@ifx{#1\undefined}
}%
\providecommand \@ifnum [1]{%
 \ifnum #1\expandafter \@firstoftwo
 \else \expandafter \@secondoftwo
 \fi
}%
\providecommand \@ifx [1]{%
 \ifx #1\expandafter \@firstoftwo
 \else \expandafter \@secondoftwo
 \fi
}%
\providecommand \natexlab [1]{#1}%
\providecommand \enquote  [1]{``#1''}%
\providecommand \bibnamefont  [1]{#1}%
\providecommand \bibfnamefont [1]{#1}%
\providecommand \citenamefont [1]{#1}%
\providecommand \href@noop [0]{\@secondoftwo}%
\providecommand \href [0]{\begingroup \@sanitize@url \@href}%
\providecommand \@href[1]{\@@startlink{#1}\@@href}%
\providecommand \@@href[1]{\endgroup#1\@@endlink}%
\providecommand \@sanitize@url [0]{\catcode `\\12\catcode `\$12\catcode
  `\&12\catcode `\#12\catcode `\^12\catcode `\_12\catcode `\%12\relax}%
\providecommand \@@startlink[1]{}%
\providecommand \@@endlink[0]{}%
\providecommand \url  [0]{\begingroup\@sanitize@url \@url }%
\providecommand \@url [1]{\endgroup\@href {#1}{\urlprefix }}%
\providecommand \urlprefix  [0]{URL }%
\providecommand \Eprint [0]{\href }%
\providecommand \doibase [0]{http://dx.doi.org/}%
\providecommand \selectlanguage [0]{\@gobble}%
\providecommand \bibinfo  [0]{\@secondoftwo}%
\providecommand \bibfield  [0]{\@secondoftwo}%
\providecommand \translation [1]{[#1]}%
\providecommand \BibitemOpen [0]{}%
\providecommand \bibitemStop [0]{}%
\providecommand \bibitemNoStop [0]{.\EOS\space}%
\providecommand \EOS [0]{\spacefactor3000\relax}%
\providecommand \BibitemShut  [1]{\csname bibitem#1\endcsname}%
\let\auto@bib@innerbib\@empty
\bibitem [{\citenamefont {Zettler}\ \emph {et~al.}(2015)\citenamefont
  {Zettler}, \citenamefont {Hauswald}, \citenamefont {Corfdir}, \citenamefont
  {Musolino}, \citenamefont {Geelhaar}, \citenamefont {Riechert}, \citenamefont
  {Brandt},\ and\ \citenamefont {Fern{\'{a}}ndez-Garrido}}]{Zettler2015a}%
  \BibitemOpen
  \bibfield  {author} {\bibinfo {author} {\bibfnamefont {J.~K.}\ \bibnamefont
  {Zettler}}, \bibinfo {author} {\bibfnamefont {C.}~\bibnamefont {Hauswald}},
  \bibinfo {author} {\bibfnamefont {P.}~\bibnamefont {Corfdir}}, \bibinfo
  {author} {\bibfnamefont {M.}~\bibnamefont {Musolino}}, \bibinfo {author}
  {\bibfnamefont {L.}~\bibnamefont {Geelhaar}}, \bibinfo {author}
  {\bibfnamefont {H.}~\bibnamefont {Riechert}}, \bibinfo {author}
  {\bibfnamefont {O.}~\bibnamefont {Brandt}}, \ and\ \bibinfo {author}
  {\bibfnamefont {S.}~\bibnamefont {Fern{\'{a}}ndez-Garrido}},\ }\href
  {\doibase 10.1021/acs.cgd.5b00690} {\bibfield  {journal} {\bibinfo  {journal}
  {Cryst. Growth Des.}\ }\textbf {\bibinfo {volume} {15}},\ \bibinfo {pages}
  {4104} (\bibinfo {year} {2015})}\BibitemShut {NoStop}%
\bibitem [{\citenamefont {Hersee}\ \emph {et~al.}(2011)\citenamefont {Hersee},
  \citenamefont {Rishinaramangalam}, \citenamefont {Fairchild}, \citenamefont
  {Zhang},\ and\ \citenamefont {Varangis}}]{Hersee2011}%
  \BibitemOpen
  \bibfield  {author} {\bibinfo {author} {\bibfnamefont {S.~D.}\ \bibnamefont
  {Hersee}}, \bibinfo {author} {\bibfnamefont {A.~K.}\ \bibnamefont
  {Rishinaramangalam}}, \bibinfo {author} {\bibfnamefont {M.~N.}\ \bibnamefont
  {Fairchild}}, \bibinfo {author} {\bibfnamefont {L.}~\bibnamefont {Zhang}}, \
  and\ \bibinfo {author} {\bibfnamefont {P.}~\bibnamefont {Varangis}},\ }\href
  {\doibase 10.1557/jmr.2011.112} {\bibfield  {journal} {\bibinfo  {journal}
  {J. Mat. Res.}\ }\textbf {\bibinfo {volume} {26}},\ \bibinfo {pages} {2293}
  (\bibinfo {year} {2011})}\BibitemShut {NoStop}%
\bibitem [{\citenamefont {Kishino}\ and\ \citenamefont
  {Ishizawa}(2015)}]{Kishino2015}%
  \BibitemOpen
  \bibfield  {author} {\bibinfo {author} {\bibfnamefont {K.}~\bibnamefont
  {Kishino}}\ and\ \bibinfo {author} {\bibfnamefont {S.}~\bibnamefont
  {Ishizawa}},\ }\href {\doibase 10.1088/0957-4484/26/22/225602} {\bibfield
  {journal} {\bibinfo  {journal} {Nanotechnology}\ }\textbf {\bibinfo {volume}
  {26}},\ \bibinfo {pages} {225602} (\bibinfo {year} {2015})}\BibitemShut
  {NoStop}%
\bibitem [{\citenamefont {Bao}\ \emph {et~al.}(2008)\citenamefont {Bao},
  \citenamefont {Bell}, \citenamefont {Capasso}, \citenamefont {Wagner},
  \citenamefont {M{\aa}rtensson}, \citenamefont {Tr\"{a}g{\aa}rdh},\ and\
  \citenamefont {Samuelson}}]{Bao2008}%
  \BibitemOpen
  \bibfield  {author} {\bibinfo {author} {\bibfnamefont {J.}~\bibnamefont
  {Bao}}, \bibinfo {author} {\bibfnamefont {D.~C.}\ \bibnamefont {Bell}},
  \bibinfo {author} {\bibfnamefont {F.}~\bibnamefont {Capasso}}, \bibinfo
  {author} {\bibfnamefont {J.~B.}\ \bibnamefont {Wagner}}, \bibinfo {author}
  {\bibfnamefont {T.}~\bibnamefont {M{\aa}rtensson}}, \bibinfo {author}
  {\bibfnamefont {J.}~\bibnamefont {Tr\"{a}g{\aa}rdh}}, \ and\ \bibinfo
  {author} {\bibfnamefont {L.}~\bibnamefont {Samuelson}},\ }\href {\doibase
  10.1021/nl072921e} {\bibfield  {journal} {\bibinfo  {journal} {Nano Lett.}\
  }\textbf {\bibinfo {volume} {8}},\ \bibinfo {pages} {836} (\bibinfo {year}
  {2008})}\BibitemShut {NoStop}%
\bibitem [{\citenamefont {Corfdir}\ \emph {et~al.}(2013)\citenamefont
  {Corfdir}, \citenamefont {{Van Hattem}}, \citenamefont {Uccelli},
  \citenamefont {Conesa-Boj}, \citenamefont {Lefebvre}, \citenamefont
  {{Fontcuberta i Morral}},\ and\ \citenamefont {Phillips}}]{Corfdir2013}%
  \BibitemOpen
  \bibfield  {author} {\bibinfo {author} {\bibfnamefont {P.}~\bibnamefont
  {Corfdir}}, \bibinfo {author} {\bibfnamefont {B.}~\bibnamefont {{Van
  Hattem}}}, \bibinfo {author} {\bibfnamefont {E.}~\bibnamefont {Uccelli}},
  \bibinfo {author} {\bibfnamefont {S.}~\bibnamefont {Conesa-Boj}}, \bibinfo
  {author} {\bibfnamefont {P.}~\bibnamefont {Lefebvre}}, \bibinfo {author}
  {\bibfnamefont {A.}~\bibnamefont {{Fontcuberta i Morral}}}, \ and\ \bibinfo
  {author} {\bibfnamefont {R.~T.}\ \bibnamefont {Phillips}},\ }\href {\doibase
  10.1021/nl4028186} {\bibfield  {journal} {\bibinfo  {journal} {Nano Lett.}\
  }\textbf {\bibinfo {volume} {13}},\ \bibinfo {pages} {5303} (\bibinfo {year}
  {2013})}\BibitemShut {NoStop}%
\bibitem [{\citenamefont {Calleja}\ \emph {et~al.}(2007)\citenamefont
  {Calleja}, \citenamefont {Risti\'{c}}, \citenamefont {Fern\'{a}ndez-Garrido},
  \citenamefont {Cerutti}, \citenamefont {S\'{a}nchez-Garc\'{\i}a},
  \citenamefont {Grandal}, \citenamefont {Trampert}, \citenamefont {Jahn},
  \citenamefont {S\'{a}nchez}, \citenamefont {Griol},\ and\ \citenamefont
  {S\'{a}nchez}}]{Calleja2007}%
  \BibitemOpen
  \bibfield  {author} {\bibinfo {author} {\bibfnamefont {E.}~\bibnamefont
  {Calleja}}, \bibinfo {author} {\bibfnamefont {J.}~\bibnamefont {Risti\'{c}}},
  \bibinfo {author} {\bibfnamefont {S.}~\bibnamefont {Fern\'{a}ndez-Garrido}},
  \bibinfo {author} {\bibfnamefont {L.}~\bibnamefont {Cerutti}}, \bibinfo
  {author} {\bibfnamefont {M.~A.}\ \bibnamefont {S\'{a}nchez-Garc\'{\i}a}},
  \bibinfo {author} {\bibfnamefont {J.}~\bibnamefont {Grandal}}, \bibinfo
  {author} {\bibfnamefont {A.}~\bibnamefont {Trampert}}, \bibinfo {author}
  {\bibfnamefont {U.}~\bibnamefont {Jahn}}, \bibinfo {author} {\bibfnamefont
  {G.}~\bibnamefont {S\'{a}nchez}}, \bibinfo {author} {\bibfnamefont
  {A.}~\bibnamefont {Griol}}, \ and\ \bibinfo {author} {\bibfnamefont
  {B.}~\bibnamefont {S\'{a}nchez}},\ }\href {\doibase 10.1002/pssb.200675628}
  {\bibfield  {journal} {\bibinfo  {journal} {Phys. Status Solidi B}\ }\textbf
  {\bibinfo {volume} {244}},\ \bibinfo {pages} {2816} (\bibinfo {year}
  {2007})}\BibitemShut {NoStop}%
\bibitem [{\citenamefont {Geelhaar}\ \emph {et~al.}(2011)\citenamefont
  {Geelhaar}, \citenamefont {Ch\`eze}, \citenamefont {Jenichen}, \citenamefont
  {Brandt}, \citenamefont {Pf{\"{u}}ller}, \citenamefont {M\"unch},
  \citenamefont {Rothemund}, \citenamefont {Reitzenstein}, \citenamefont
  {Forchel}, \citenamefont {Kehagias}, \citenamefont {Komninou}, \citenamefont
  {Dimitrakopulos}, \citenamefont {Karakostas}, \citenamefont {Lari},
  \citenamefont {Chalker}, \citenamefont {Gass},\ and\ \citenamefont
  {Riechert}}]{Geelhaar2011}%
  \BibitemOpen
  \bibfield  {author} {\bibinfo {author} {\bibfnamefont {L.}~\bibnamefont
  {Geelhaar}}, \bibinfo {author} {\bibfnamefont {C.}~\bibnamefont {Ch\`eze}},
  \bibinfo {author} {\bibfnamefont {B.}~\bibnamefont {Jenichen}}, \bibinfo
  {author} {\bibfnamefont {O.}~\bibnamefont {Brandt}}, \bibinfo {author}
  {\bibfnamefont {C.}~\bibnamefont {Pf{\"{u}}ller}}, \bibinfo {author}
  {\bibfnamefont {S.}~\bibnamefont {M\"unch}}, \bibinfo {author} {\bibfnamefont
  {R.}~\bibnamefont {Rothemund}}, \bibinfo {author} {\bibfnamefont
  {S.}~\bibnamefont {Reitzenstein}}, \bibinfo {author} {\bibfnamefont
  {A.}~\bibnamefont {Forchel}}, \bibinfo {author} {\bibfnamefont
  {T.}~\bibnamefont {Kehagias}}, \bibinfo {author} {\bibfnamefont
  {P.}~\bibnamefont {Komninou}}, \bibinfo {author} {\bibfnamefont {G.~P.}\
  \bibnamefont {Dimitrakopulos}}, \bibinfo {author} {\bibfnamefont
  {T.}~\bibnamefont {Karakostas}}, \bibinfo {author} {\bibfnamefont
  {L.}~\bibnamefont {Lari}}, \bibinfo {author} {\bibfnamefont {P.~R.}\
  \bibnamefont {Chalker}}, \bibinfo {author} {\bibfnamefont {M.~H.}\
  \bibnamefont {Gass}}, \ and\ \bibinfo {author} {\bibfnamefont
  {H.}~\bibnamefont {Riechert}},\ }\href {\doibase 10.1109/JSTQE.2010.2098396}
  {\bibfield  {journal} {\bibinfo  {journal} {IEEE J. Sel. Top. Quantum
  Electro.}\ }\textbf {\bibinfo {volume} {17}},\ \bibinfo {pages} {878}
  (\bibinfo {year} {2011})}\BibitemShut {NoStop}%
\bibitem [{\citenamefont {Corfdir}\ \emph {et~al.}(2014)\citenamefont
  {Corfdir}, \citenamefont {Hauswald}, \citenamefont {Zettler}, \citenamefont
  {Flissikowski}, \citenamefont {L\"{a}hnemann}, \citenamefont
  {Fern\'{a}ndez-Garrido}, \citenamefont {Geelhaar}, \citenamefont {Grahn},\
  and\ \citenamefont {Brandt}}]{Corfdir2014b}%
  \BibitemOpen
  \bibfield  {author} {\bibinfo {author} {\bibfnamefont {P.}~\bibnamefont
  {Corfdir}}, \bibinfo {author} {\bibfnamefont {C.}~\bibnamefont {Hauswald}},
  \bibinfo {author} {\bibfnamefont {J.~K.}\ \bibnamefont {Zettler}}, \bibinfo
  {author} {\bibfnamefont {T.}~\bibnamefont {Flissikowski}}, \bibinfo {author}
  {\bibfnamefont {J.}~\bibnamefont {L\"{a}hnemann}}, \bibinfo {author}
  {\bibfnamefont {S.}~\bibnamefont {Fern\'{a}ndez-Garrido}}, \bibinfo {author}
  {\bibfnamefont {L.}~\bibnamefont {Geelhaar}}, \bibinfo {author}
  {\bibfnamefont {H.~T.}\ \bibnamefont {Grahn}}, \ and\ \bibinfo {author}
  {\bibfnamefont {O.}~\bibnamefont {Brandt}},\ }\href {\doibase
  10.1103/PhysRevB.90.195309} {\bibfield  {journal} {\bibinfo  {journal} {Phys.
  Rev. B}\ }\textbf {\bibinfo {volume} {90}},\ \bibinfo {pages} {195309}
  (\bibinfo {year} {2014})}\BibitemShut {NoStop}%
\bibitem [{\citenamefont {Korona}\ \emph {et~al.}(2014)\citenamefont {Korona},
  \citenamefont {Reszka}, \citenamefont {Sobanska}, \citenamefont {Perkowska},
  \citenamefont {Wysmołek}, \citenamefont {Klosek},\ and\ \citenamefont
  {Zytkiewicz}}]{Korona2014}%
  \BibitemOpen
  \bibfield  {author} {\bibinfo {author} {\bibfnamefont {K.~P.}\ \bibnamefont
  {Korona}}, \bibinfo {author} {\bibfnamefont {A.}~\bibnamefont {Reszka}},
  \bibinfo {author} {\bibfnamefont {M.}~\bibnamefont {Sobanska}}, \bibinfo
  {author} {\bibfnamefont {P.}~\bibnamefont {Perkowska}}, \bibinfo {author}
  {\bibfnamefont {A.}~\bibnamefont {Wysmołek}}, \bibinfo {author}
  {\bibfnamefont {K.}~\bibnamefont {Klosek}}, \ and\ \bibinfo {author}
  {\bibfnamefont {Z.~R.}\ \bibnamefont {Zytkiewicz}},\ }\href {\doibase
  10.1016/j.jlumin.2014.06.061} {\bibfield  {journal} {\bibinfo  {journal} {J.
  Lumin.}\ }\textbf {\bibinfo {volume} {155}},\ \bibinfo {pages} {293}
  (\bibinfo {year} {2014})}\BibitemShut {NoStop}%
\bibitem [{\citenamefont {Stampfl}\ and\ \citenamefont {{Van de
  Walle}}(1998)}]{Stampfl1998}%
  \BibitemOpen
  \bibfield  {author} {\bibinfo {author} {\bibfnamefont {C.}~\bibnamefont
  {Stampfl}}\ and\ \bibinfo {author} {\bibfnamefont {C.~G.}\ \bibnamefont {{Van
  de Walle}}},\ }\href {\doibase 10.1103/PhysRevB.57.R15052} {\bibfield
  {journal} {\bibinfo  {journal} {Phys. Rev. B}\ }\textbf {\bibinfo {volume}
  {57}},\ \bibinfo {pages} {R15052} (\bibinfo {year} {1998})}\BibitemShut
  {NoStop}%
\bibitem [{\citenamefont {Akopian}\ \emph {et~al.}(2010)\citenamefont
  {Akopian}, \citenamefont {Patriarche}, \citenamefont {Liu}, \citenamefont
  {Harmand},\ and\ \citenamefont {Zwiller}}]{Akopian2010}%
  \BibitemOpen
  \bibfield  {author} {\bibinfo {author} {\bibfnamefont {N.}~\bibnamefont
  {Akopian}}, \bibinfo {author} {\bibfnamefont {G.}~\bibnamefont {Patriarche}},
  \bibinfo {author} {\bibfnamefont {L.}~\bibnamefont {Liu}}, \bibinfo {author}
  {\bibfnamefont {J.-C.}\ \bibnamefont {Harmand}}, \ and\ \bibinfo {author}
  {\bibfnamefont {V.}~\bibnamefont {Zwiller}},\ }\href {\doibase
  10.1021/nl903534n} {\bibfield  {journal} {\bibinfo  {journal} {Nano Lett.}\
  }\textbf {\bibinfo {volume} {10}},\ \bibinfo {pages} {1198} (\bibinfo {year}
  {2010})}\BibitemShut {NoStop}%
\bibitem [{\citenamefont {Zettler}\ \emph {et~al.}(2016)\citenamefont
  {Zettler}, \citenamefont {Corfdir}, \citenamefont {Hauswald}, \citenamefont
  {Luna}, \citenamefont {Jahn}, \citenamefont {Flissikowski}, \citenamefont
  {Schmidt}, \citenamefont {Ronning}, \citenamefont {Trampert}, \citenamefont
  {Geelhaar}, \citenamefont {Grahn}, \citenamefont {Brandt},\ and\
  \citenamefont {Fern{\'{a}}ndez-Garrido}}]{Zettler2016}%
  \BibitemOpen
  \bibfield  {author} {\bibinfo {author} {\bibfnamefont {J.~K.}\ \bibnamefont
  {Zettler}}, \bibinfo {author} {\bibfnamefont {P.}~\bibnamefont {Corfdir}},
  \bibinfo {author} {\bibfnamefont {C.}~\bibnamefont {Hauswald}}, \bibinfo
  {author} {\bibfnamefont {E.}~\bibnamefont {Luna}}, \bibinfo {author}
  {\bibfnamefont {U.}~\bibnamefont {Jahn}}, \bibinfo {author} {\bibfnamefont
  {T.}~\bibnamefont {Flissikowski}}, \bibinfo {author} {\bibfnamefont
  {E.}~\bibnamefont {Schmidt}}, \bibinfo {author} {\bibfnamefont
  {C.}~\bibnamefont {Ronning}}, \bibinfo {author} {\bibfnamefont
  {A.}~\bibnamefont {Trampert}}, \bibinfo {author} {\bibfnamefont
  {L.}~\bibnamefont {Geelhaar}}, \bibinfo {author} {\bibfnamefont {H.~T.}\
  \bibnamefont {Grahn}}, \bibinfo {author} {\bibfnamefont {O.}~\bibnamefont
  {Brandt}}, \ and\ \bibinfo {author} {\bibfnamefont {S.}~\bibnamefont
  {Fern{\'{a}}ndez-Garrido}},\ }\href {\doibase 10.1021/acs.nanolett.5b03931}
  {\bibfield  {journal} {\bibinfo  {journal} {Nano Lett.}\ }\textbf {\bibinfo
  {volume} {16}},\ \bibinfo {pages} {973} (\bibinfo {year} {2016})}\BibitemShut
  {NoStop}%
\bibitem [{\citenamefont {Loitsch}\ \emph {et~al.}(2015)\citenamefont
  {Loitsch}, \citenamefont {Rudolph}, \citenamefont {Mork\"{o}tter},
  \citenamefont {D\"{o}blinger}, \citenamefont {Grimaldi}, \citenamefont
  {Hanschke}, \citenamefont {Matich}, \citenamefont {Parzinger}, \citenamefont
  {Wurstbauer}, \citenamefont {Abstreiter}, \citenamefont {Finley},\ and\
  \citenamefont {Koblm\"{u}ller}}]{Loitsch2015}%
  \BibitemOpen
  \bibfield  {author} {\bibinfo {author} {\bibfnamefont {B.}~\bibnamefont
  {Loitsch}}, \bibinfo {author} {\bibfnamefont {D.}~\bibnamefont {Rudolph}},
  \bibinfo {author} {\bibfnamefont {S.}~\bibnamefont {Mork\"{o}tter}}, \bibinfo
  {author} {\bibfnamefont {M.}~\bibnamefont {D\"{o}blinger}}, \bibinfo {author}
  {\bibfnamefont {G.}~\bibnamefont {Grimaldi}}, \bibinfo {author}
  {\bibfnamefont {L.}~\bibnamefont {Hanschke}}, \bibinfo {author}
  {\bibfnamefont {S.}~\bibnamefont {Matich}}, \bibinfo {author} {\bibfnamefont
  {E.}~\bibnamefont {Parzinger}}, \bibinfo {author} {\bibfnamefont
  {U.}~\bibnamefont {Wurstbauer}}, \bibinfo {author} {\bibfnamefont
  {G.}~\bibnamefont {Abstreiter}}, \bibinfo {author} {\bibfnamefont {J.~J.}\
  \bibnamefont {Finley}}, \ and\ \bibinfo {author} {\bibfnamefont
  {G.}~\bibnamefont {Koblm\"{u}ller}},\ }\href {\doibase
  10.1002/adma.201404900} {\bibfield  {journal} {\bibinfo  {journal} {Adv.
  Mater.}\ ,\ \bibinfo {pages} {2195}} (\bibinfo {year} {2015})}\BibitemShut
  {NoStop}%
\bibitem [{\citenamefont {Keldysh}(1979)}]{Keldysh1979}%
  \BibitemOpen
  \bibfield  {author} {\bibinfo {author} {\bibfnamefont {L.~V.}\ \bibnamefont
  {Keldysh}},\ }\href
  {http://www.jetpletters.ac.ru/ps/1458/article_22207.shtml} {\bibfield
  {journal} {\bibinfo  {journal} {Pis´ma Zh. Eksp. Teor. Fiz.}\ }\textbf
  {\bibinfo {volume} {29}},\ \bibinfo {pages} {716} (\bibinfo {year} {1979})},\
  \bibinfo {note} {[JETP Lett. \textbf{29}, 658 (1979)]}\BibitemShut {NoStop}%
\bibitem [{\citenamefont {Kumagai}\ and\ \citenamefont
  {Takagahara}(1989)}]{Kumagai1989}%
  \BibitemOpen
  \bibfield  {author} {\bibinfo {author} {\bibfnamefont {M.}~\bibnamefont
  {Kumagai}}\ and\ \bibinfo {author} {\bibfnamefont {T.}~\bibnamefont
  {Takagahara}},\ }\href {\doibase 10.1103/PhysRevB.40.12359} {\bibfield
  {journal} {\bibinfo  {journal} {Phys. Rev. B}\ }\textbf {\bibinfo {volume}
  {40}},\ \bibinfo {pages} {12359} (\bibinfo {year} {1989})}\BibitemShut
  {NoStop}%
\bibitem [{\citenamefont {Consonni}\ \emph {et~al.}(2009)\citenamefont
  {Consonni}, \citenamefont {Knelangen}, \citenamefont {Jahn}, \citenamefont
  {Trampert}, \citenamefont {Geelhaar},\ and\ \citenamefont
  {Riechert}}]{Consonni2009}%
  \BibitemOpen
  \bibfield  {author} {\bibinfo {author} {\bibfnamefont {V.}~\bibnamefont
  {Consonni}}, \bibinfo {author} {\bibfnamefont {M.}~\bibnamefont {Knelangen}},
  \bibinfo {author} {\bibfnamefont {U.}~\bibnamefont {Jahn}}, \bibinfo {author}
  {\bibfnamefont {A.}~\bibnamefont {Trampert}}, \bibinfo {author}
  {\bibfnamefont {L.}~\bibnamefont {Geelhaar}}, \ and\ \bibinfo {author}
  {\bibfnamefont {H.}~\bibnamefont {Riechert}},\ }\href {\doibase
  10.1063/1.3275793} {\bibfield  {journal} {\bibinfo  {journal} {Appl. Phys.
  Lett.}\ }\textbf {\bibinfo {volume} {95}},\ \bibinfo {pages} {241910}
  (\bibinfo {year} {2009})}\BibitemShut {NoStop}%
\bibitem [{\citenamefont {Graham}\ \emph {et~al.}(2013)\citenamefont {Graham},
  \citenamefont {Corfdir}, \citenamefont {Heiss}, \citenamefont {Conesa-Boj},
  \citenamefont {Uccelli}, \citenamefont {Fontcuberta~i Morral},\ and\
  \citenamefont {Phillips}}]{Graham2013}%
  \BibitemOpen
  \bibfield  {author} {\bibinfo {author} {\bibfnamefont {A.~M.}\ \bibnamefont
  {Graham}}, \bibinfo {author} {\bibfnamefont {P.}~\bibnamefont {Corfdir}},
  \bibinfo {author} {\bibfnamefont {M.}~\bibnamefont {Heiss}}, \bibinfo
  {author} {\bibfnamefont {S.}~\bibnamefont {Conesa-Boj}}, \bibinfo {author}
  {\bibfnamefont {E.}~\bibnamefont {Uccelli}}, \bibinfo {author} {\bibfnamefont
  {A.}~\bibnamefont {Fontcuberta~i Morral}}, \ and\ \bibinfo {author}
  {\bibfnamefont {R.~T.}\ \bibnamefont {Phillips}},\ }\href {\doibase
  10.1103/PhysRevB.87.125304} {\bibfield  {journal} {\bibinfo  {journal} {Phys.
  Rev. B}\ }\textbf {\bibinfo {volume} {87}},\ \bibinfo {pages} {125304}
  (\bibinfo {year} {2013})}\BibitemShut {NoStop}%
\bibitem [{\citenamefont {Rudolph}\ \emph {et~al.}(2013)\citenamefont
  {Rudolph}, \citenamefont {Schweickert}, \citenamefont {Mork\"{o}tter},
  \citenamefont {Hanschke}, \citenamefont {Hertenberger}, \citenamefont
  {Bichler}, \citenamefont {Koblm\"{u}ller}, \citenamefont {Abstreiter},\ and\
  \citenamefont {Finley}}]{Rudolph2013}%
  \BibitemOpen
  \bibfield  {author} {\bibinfo {author} {\bibfnamefont {D.}~\bibnamefont
  {Rudolph}}, \bibinfo {author} {\bibfnamefont {L.}~\bibnamefont
  {Schweickert}}, \bibinfo {author} {\bibfnamefont {S.}~\bibnamefont
  {Mork\"{o}tter}}, \bibinfo {author} {\bibfnamefont {L.}~\bibnamefont
  {Hanschke}}, \bibinfo {author} {\bibfnamefont {S.}~\bibnamefont
  {Hertenberger}}, \bibinfo {author} {\bibfnamefont {M.}~\bibnamefont
  {Bichler}}, \bibinfo {author} {\bibfnamefont {G.}~\bibnamefont
  {Koblm\"{u}ller}}, \bibinfo {author} {\bibfnamefont {G.}~\bibnamefont
  {Abstreiter}}, \ and\ \bibinfo {author} {\bibfnamefont {J.~J.}\ \bibnamefont
  {Finley}},\ }\href {\doibase 10.1088/1367-2630/15/11/113032} {\bibfield
  {journal} {\bibinfo  {journal} {New J. Phys.}\ }\textbf {\bibinfo {volume}
  {15}},\ \bibinfo {pages} {113032} (\bibinfo {year} {2013})}\BibitemShut
  {NoStop}%
\bibitem [{\citenamefont {Hauswald}\ \emph {et~al.}(2014)\citenamefont
  {Hauswald}, \citenamefont {Corfdir}, \citenamefont {Zettler}, \citenamefont
  {Kaganer}, \citenamefont {Sabelfeld}, \citenamefont
  {Fern{\'{a}}ndez-Garrido}, \citenamefont {Flissikowski}, \citenamefont
  {Consonni}, \citenamefont {Gotschke}, \citenamefont {Grahn}, \citenamefont
  {Geelhaar},\ and\ \citenamefont {Brandt}}]{Hauswald2014}%
  \BibitemOpen
  \bibfield  {author} {\bibinfo {author} {\bibfnamefont {C.}~\bibnamefont
  {Hauswald}}, \bibinfo {author} {\bibfnamefont {P.}~\bibnamefont {Corfdir}},
  \bibinfo {author} {\bibfnamefont {J.~K.}\ \bibnamefont {Zettler}}, \bibinfo
  {author} {\bibfnamefont {V.~M.}\ \bibnamefont {Kaganer}}, \bibinfo {author}
  {\bibfnamefont {K.~K.}\ \bibnamefont {Sabelfeld}}, \bibinfo {author}
  {\bibfnamefont {S.}~\bibnamefont {Fern{\'{a}}ndez-Garrido}}, \bibinfo
  {author} {\bibfnamefont {T.}~\bibnamefont {Flissikowski}}, \bibinfo {author}
  {\bibfnamefont {V.}~\bibnamefont {Consonni}}, \bibinfo {author}
  {\bibfnamefont {T.}~\bibnamefont {Gotschke}}, \bibinfo {author}
  {\bibfnamefont {H.~T.}\ \bibnamefont {Grahn}}, \bibinfo {author}
  {\bibfnamefont {L.}~\bibnamefont {Geelhaar}}, \ and\ \bibinfo {author}
  {\bibfnamefont {O.}~\bibnamefont {Brandt}},\ }\href {\doibase
  10.1103/PhysRevB.90.165304} {\bibfield  {journal} {\bibinfo  {journal} {Phys.
  Rev. B}\ }\textbf {\bibinfo {volume} {90}},\ \bibinfo {pages} {165304}
  (\bibinfo {year} {2014})}\BibitemShut {NoStop}%
\bibitem [{\citenamefont {Hauswald}\ \emph {et~al.}(2013)\citenamefont
  {Hauswald}, \citenamefont {Flissikowski}, \citenamefont {Gotschke},
  \citenamefont {Calarco}, \citenamefont {Geelhaar}, \citenamefont {Grahn},\
  and\ \citenamefont {Brandt}}]{Hauswald2013}%
  \BibitemOpen
  \bibfield  {author} {\bibinfo {author} {\bibfnamefont {C.}~\bibnamefont
  {Hauswald}}, \bibinfo {author} {\bibfnamefont {T.}~\bibnamefont
  {Flissikowski}}, \bibinfo {author} {\bibfnamefont {T.}~\bibnamefont
  {Gotschke}}, \bibinfo {author} {\bibfnamefont {R.}~\bibnamefont {Calarco}},
  \bibinfo {author} {\bibfnamefont {L.}~\bibnamefont {Geelhaar}}, \bibinfo
  {author} {\bibfnamefont {H.~T.}\ \bibnamefont {Grahn}}, \ and\ \bibinfo
  {author} {\bibfnamefont {O.}~\bibnamefont {Brandt}},\ }\href {\doibase
  10.1103/PhysRevB.88.075312} {\bibfield  {journal} {\bibinfo  {journal} {Phys.
  Rev. B}\ }\textbf {\bibinfo {volume} {88}},\ \bibinfo {pages} {075312}
  (\bibinfo {year} {2013})}\BibitemShut {NoStop}%
\bibitem [{\citenamefont {Sun}\ \emph {et~al.}(2002)\citenamefont {Sun},
  \citenamefont {Brandt}, \citenamefont {Jahn}, \citenamefont {Liu},
  \citenamefont {Trampert}, \citenamefont {Cronenberg}, \citenamefont {Dhar},\
  and\ \citenamefont {Ploog}}]{Sun2002}%
  \BibitemOpen
  \bibfield  {author} {\bibinfo {author} {\bibfnamefont {Y.~J.}\ \bibnamefont
  {Sun}}, \bibinfo {author} {\bibfnamefont {O.}~\bibnamefont {Brandt}},
  \bibinfo {author} {\bibfnamefont {U.}~\bibnamefont {Jahn}}, \bibinfo {author}
  {\bibfnamefont {T.~Y.}\ \bibnamefont {Liu}}, \bibinfo {author} {\bibfnamefont
  {A.}~\bibnamefont {Trampert}}, \bibinfo {author} {\bibfnamefont
  {S.}~\bibnamefont {Cronenberg}}, \bibinfo {author} {\bibfnamefont
  {S.}~\bibnamefont {Dhar}}, \ and\ \bibinfo {author} {\bibfnamefont {K.~H.}\
  \bibnamefont {Ploog}},\ }\href {\doibase 10.1063/1.1513874} {\bibfield
  {journal} {\bibinfo  {journal} {J. Appl. Phys.}\ }\textbf {\bibinfo {volume}
  {92}},\ \bibinfo {pages} {5714} (\bibinfo {year} {2002})}\BibitemShut
  {NoStop}%
\bibitem [{\citenamefont {L{\"{a}}hnemann}\ \emph {et~al.}(2012)\citenamefont
  {L{\"{a}}hnemann}, \citenamefont {Brandt}, \citenamefont {Jahn},
  \citenamefont {Pf{\"{u}}ller}, \citenamefont {Roder}, \citenamefont {Dogan},
  \citenamefont {Grosse}, \citenamefont {Belabbes}, \citenamefont {Bechstedt},
  \citenamefont {Trampert},\ and\ \citenamefont {Geelhaar}}]{Laehnemann2012}%
  \BibitemOpen
  \bibfield  {author} {\bibinfo {author} {\bibfnamefont {J.}~\bibnamefont
  {L{\"{a}}hnemann}}, \bibinfo {author} {\bibfnamefont {O.}~\bibnamefont
  {Brandt}}, \bibinfo {author} {\bibfnamefont {U.}~\bibnamefont {Jahn}},
  \bibinfo {author} {\bibfnamefont {C.}~\bibnamefont {Pf{\"{u}}ller}}, \bibinfo
  {author} {\bibfnamefont {C.}~\bibnamefont {Roder}}, \bibinfo {author}
  {\bibfnamefont {P.}~\bibnamefont {Dogan}}, \bibinfo {author} {\bibfnamefont
  {F.}~\bibnamefont {Grosse}}, \bibinfo {author} {\bibfnamefont
  {A.}~\bibnamefont {Belabbes}}, \bibinfo {author} {\bibfnamefont
  {F.}~\bibnamefont {Bechstedt}}, \bibinfo {author} {\bibfnamefont
  {A.}~\bibnamefont {Trampert}}, \ and\ \bibinfo {author} {\bibfnamefont
  {L.}~\bibnamefont {Geelhaar}},\ }\href {\doibase 10.1103/PhysRevB.86.081302}
  {\bibfield  {journal} {\bibinfo  {journal} {Phys. Rev. B}\ }\textbf {\bibinfo
  {volume} {86}},\ \bibinfo {pages} {081302} (\bibinfo {year}
  {2012})}\BibitemShut {NoStop}%
\bibitem [{\citenamefont {Corfdir}\ and\ \citenamefont
  {Lefebvre}(2012)}]{Corfdir2012}%
  \BibitemOpen
  \bibfield  {author} {\bibinfo {author} {\bibfnamefont {P.}~\bibnamefont
  {Corfdir}}\ and\ \bibinfo {author} {\bibfnamefont {P.}~\bibnamefont
  {Lefebvre}},\ }\href {\doibase 10.1063/1.4749789} {\bibfield  {journal}
  {\bibinfo  {journal} {J. Appl. Phys.}\ }\textbf {\bibinfo {volume} {112}},\
  \bibinfo {pages} {053512} (\bibinfo {year} {2012})}\BibitemShut {NoStop}%
\bibitem [{\citenamefont {Reale}\ \emph {et~al.}(2003)\citenamefont {Reale},
  \citenamefont {Massari}, \citenamefont {Di~Carlo}, \citenamefont {Lugli},
  \citenamefont {Vinattieri}, \citenamefont {Alderighi}, \citenamefont
  {Colocci}, \citenamefont {Semond}, \citenamefont {Grandjean},\ and\
  \citenamefont {Massies}}]{Reale2003}%
  \BibitemOpen
  \bibfield  {author} {\bibinfo {author} {\bibfnamefont {A.}~\bibnamefont
  {Reale}}, \bibinfo {author} {\bibfnamefont {G.}~\bibnamefont {Massari}},
  \bibinfo {author} {\bibfnamefont {A.}~\bibnamefont {Di~Carlo}}, \bibinfo
  {author} {\bibfnamefont {P.}~\bibnamefont {Lugli}}, \bibinfo {author}
  {\bibfnamefont {A.}~\bibnamefont {Vinattieri}}, \bibinfo {author}
  {\bibfnamefont {D.}~\bibnamefont {Alderighi}}, \bibinfo {author}
  {\bibfnamefont {M.}~\bibnamefont {Colocci}}, \bibinfo {author} {\bibfnamefont
  {F.}~\bibnamefont {Semond}}, \bibinfo {author} {\bibfnamefont
  {N.}~\bibnamefont {Grandjean}}, \ and\ \bibinfo {author} {\bibfnamefont
  {J.}~\bibnamefont {Massies}},\ }\href {\doibase
  http://dx.doi.org/10.1063/1.1527989} {\bibfield  {journal} {\bibinfo
  {journal} {J. Appl. Phys.}\ }\textbf {\bibinfo {volume} {93}},\ \bibinfo
  {pages} {400} (\bibinfo {year} {2003})}\BibitemShut {NoStop}%
\bibitem [{\citenamefont {Bretagnon}\ \emph {et~al.}(2006)\citenamefont
  {Bretagnon}, \citenamefont {Lefebvre}, \citenamefont {Valvin}, \citenamefont
  {Bardoux}, \citenamefont {Guillet}, \citenamefont {Taliercio}, \citenamefont
  {Gil}, \citenamefont {Grandjean}, \citenamefont {Semond}, \citenamefont
  {Damilano}, \citenamefont {Dussaigne},\ and\ \citenamefont
  {Massies}}]{Bretagnon2006}%
  \BibitemOpen
  \bibfield  {author} {\bibinfo {author} {\bibfnamefont {T.}~\bibnamefont
  {Bretagnon}}, \bibinfo {author} {\bibfnamefont {P.}~\bibnamefont {Lefebvre}},
  \bibinfo {author} {\bibfnamefont {P.}~\bibnamefont {Valvin}}, \bibinfo
  {author} {\bibfnamefont {R.}~\bibnamefont {Bardoux}}, \bibinfo {author}
  {\bibfnamefont {T.}~\bibnamefont {Guillet}}, \bibinfo {author} {\bibfnamefont
  {T.}~\bibnamefont {Taliercio}}, \bibinfo {author} {\bibfnamefont
  {B.}~\bibnamefont {Gil}}, \bibinfo {author} {\bibfnamefont {N.}~\bibnamefont
  {Grandjean}}, \bibinfo {author} {\bibfnamefont {F.}~\bibnamefont {Semond}},
  \bibinfo {author} {\bibfnamefont {B.}~\bibnamefont {Damilano}}, \bibinfo
  {author} {\bibfnamefont {A.}~\bibnamefont {Dussaigne}}, \ and\ \bibinfo
  {author} {\bibfnamefont {J.}~\bibnamefont {Massies}},\ }\href {\doibase
  10.1103/PhysRevB.73.113304} {\bibfield  {journal} {\bibinfo  {journal} {Phys.
  Rev. B}\ }\textbf {\bibinfo {volume} {73}},\ \bibinfo {pages} {113304}
  (\bibinfo {year} {2006})}\BibitemShut {NoStop}%
\bibitem [{\citenamefont {Kavokin}(1994)}]{Kavokin1994}%
  \BibitemOpen
  \bibfield  {author} {\bibinfo {author} {\bibfnamefont {A.~V.}\ \bibnamefont
  {Kavokin}},\ }\href {\doibase 10.1103/PhysRevB.50.8000} {\bibfield  {journal}
  {\bibinfo  {journal} {Phys. Rev. B}\ }\textbf {\bibinfo {volume} {50}},\
  \bibinfo {pages} {8000} (\bibinfo {year} {1994})}\BibitemShut {NoStop}%
\bibitem [{\citenamefont {Bellessa}\ \emph {et~al.}(1998)\citenamefont
  {Bellessa}, \citenamefont {Voliotis}, \citenamefont {Grousson}, \citenamefont
  {Wang}, \citenamefont {Ogura},\ and\ \citenamefont
  {Matsuhata}}]{Bellessa1998}%
  \BibitemOpen
  \bibfield  {author} {\bibinfo {author} {\bibfnamefont {J.}~\bibnamefont
  {Bellessa}}, \bibinfo {author} {\bibfnamefont {V.}~\bibnamefont {Voliotis}},
  \bibinfo {author} {\bibfnamefont {R.}~\bibnamefont {Grousson}}, \bibinfo
  {author} {\bibfnamefont {X.~L.}\ \bibnamefont {Wang}}, \bibinfo {author}
  {\bibfnamefont {M.}~\bibnamefont {Ogura}}, \ and\ \bibinfo {author}
  {\bibfnamefont {H.}~\bibnamefont {Matsuhata}},\ }\href {\doibase
  10.1103/PhysRevB.58.9933} {\bibfield  {journal} {\bibinfo  {journal} {Phys.
  Rev. B}\ }\textbf {\bibinfo {volume} {58}},\ \bibinfo {pages} {9933}
  (\bibinfo {year} {1998})}\BibitemShut {NoStop}%
\bibitem [{\citenamefont {Corfdir}\ \emph {et~al.}(2011)\citenamefont
  {Corfdir}, \citenamefont {Levrat}, \citenamefont {Dussaigne}, \citenamefont
  {Lefebvre}, \citenamefont {Teisseyre}, \citenamefont {Grzegory},
  \citenamefont {Suski}, \citenamefont {Gani\`{e}re}, \citenamefont
  {Grandjean},\ and\ \citenamefont {Deveaud-Pl\'{e}dran}}]{Corfdir2011}%
  \BibitemOpen
  \bibfield  {author} {\bibinfo {author} {\bibfnamefont {P.}~\bibnamefont
  {Corfdir}}, \bibinfo {author} {\bibfnamefont {J.}~\bibnamefont {Levrat}},
  \bibinfo {author} {\bibfnamefont {A.}~\bibnamefont {Dussaigne}}, \bibinfo
  {author} {\bibfnamefont {P.}~\bibnamefont {Lefebvre}}, \bibinfo {author}
  {\bibfnamefont {H.}~\bibnamefont {Teisseyre}}, \bibinfo {author}
  {\bibfnamefont {I.}~\bibnamefont {Grzegory}}, \bibinfo {author}
  {\bibfnamefont {T.}~\bibnamefont {Suski}}, \bibinfo {author} {\bibfnamefont
  {J.-D.}\ \bibnamefont {Gani\`{e}re}}, \bibinfo {author} {\bibfnamefont
  {N.}~\bibnamefont {Grandjean}}, \ and\ \bibinfo {author} {\bibfnamefont
  {B.}~\bibnamefont {Deveaud-Pl\'{e}dran}},\ }\href {\doibase
  10.1103/PhysRevB.83.245326} {\bibfield  {journal} {\bibinfo  {journal} {Phys.
  Rev. B}\ }\textbf {\bibinfo {volume} {83}},\ \bibinfo {pages} {245326}
  (\bibinfo {year} {2011})}\BibitemShut {NoStop}%
\bibitem [{\citenamefont {Humphrey}\ \emph {et~al.}(1996)\citenamefont
  {Humphrey}, \citenamefont {Dalke},\ and\ \citenamefont
  {Schulten}}]{Humphrey1996}%
  \BibitemOpen
  \bibfield  {author} {\bibinfo {author} {\bibfnamefont {W.}~\bibnamefont
  {Humphrey}}, \bibinfo {author} {\bibfnamefont {A.}~\bibnamefont {Dalke}}, \
  and\ \bibinfo {author} {\bibfnamefont {K.}~\bibnamefont {Schulten}},\
  }\href@noop {} {\bibfield  {journal} {\bibinfo  {journal} {J. Mol. Graph.}\
  }\textbf {\bibinfo {volume} {14}},\ \bibinfo {pages} {33} (\bibinfo {year}
  {1996})}\BibitemShut {NoStop}%
\bibitem [{\citenamefont {Brandt}\ \emph {et~al.}(1992)\citenamefont {Brandt},
  \citenamefont {La~Rocca}, \citenamefont {Heberle}, \citenamefont {Ruiz},\
  and\ \citenamefont {Ploog}}]{Brandt1992}%
  \BibitemOpen
  \bibfield  {author} {\bibinfo {author} {\bibfnamefont {O.}~\bibnamefont
  {Brandt}}, \bibinfo {author} {\bibfnamefont {G.~C.}\ \bibnamefont
  {La~Rocca}}, \bibinfo {author} {\bibfnamefont {A.}~\bibnamefont {Heberle}},
  \bibinfo {author} {\bibfnamefont {A.}~\bibnamefont {Ruiz}}, \ and\ \bibinfo
  {author} {\bibfnamefont {K.}~\bibnamefont {Ploog}},\ }\href {\doibase
  10.1103/PhysRevB.45.3803} {\bibfield  {journal} {\bibinfo  {journal} {Phys.
  Rev. B}\ }\textbf {\bibinfo {volume} {45}},\ \bibinfo {pages} {3803}
  (\bibinfo {year} {1992})}\BibitemShut {NoStop}%
\bibitem [{\citenamefont {Marquardt}\ \emph {et~al.}(2014)\citenamefont
  {Marquardt}, \citenamefont {Boeck}, \citenamefont {Freysoldt}, \citenamefont
  {Hickel}, \citenamefont {Schulz}, \citenamefont {Neugebauer},\ and\
  \citenamefont {O’Reilly}}]{Marquardt2014}%
  \BibitemOpen
  \bibfield  {author} {\bibinfo {author} {\bibfnamefont {O.}~\bibnamefont
  {Marquardt}}, \bibinfo {author} {\bibfnamefont {S.}~\bibnamefont {Boeck}},
  \bibinfo {author} {\bibfnamefont {C.}~\bibnamefont {Freysoldt}}, \bibinfo
  {author} {\bibfnamefont {T.}~\bibnamefont {Hickel}}, \bibinfo {author}
  {\bibfnamefont {S.}~\bibnamefont {Schulz}}, \bibinfo {author} {\bibfnamefont
  {J.}~\bibnamefont {Neugebauer}}, \ and\ \bibinfo {author} {\bibfnamefont
  {E.~P.}\ \bibnamefont {O’Reilly}},\ }\href {\doibase
  http://dx.doi.org/10.1016/j.commatsci.2014.06.047} {\bibfield  {journal}
  {\bibinfo  {journal} {Comp. Mat. Sci.}\ }\textbf {\bibinfo {volume} {95}},\
  \bibinfo {pages} {280 } (\bibinfo {year} {2014})}\BibitemShut {NoStop}%
\bibitem [{\citenamefont {Rashba}\ and\ \citenamefont
  {Gurgenishvili}(1962)}]{Rashba1962}%
  \BibitemOpen
  \bibfield  {author} {\bibinfo {author} {\bibfnamefont {E.~I.}\ \bibnamefont
  {Rashba}}\ and\ \bibinfo {author} {\bibfnamefont {G.~E.}\ \bibnamefont
  {Gurgenishvili}},\ }\href@noop {} {\bibfield  {journal} {\bibinfo  {journal}
  {Fiz. Tverd. Tela}\ }\textbf {\bibinfo {volume} {4}},\ \bibinfo {pages}
  {1029} (\bibinfo {year} {1962})},\ \bibinfo {note} {[Sov. Phys. - Solid State
  \textbf{4}, 759 (1962)]}\BibitemShut {NoStop}%
\end{thebibliography}%

\end{document}